\documentclass[preprint]{vgtc}               

\ifpdf
  \pdfoutput=1\relax                   
  \usepackage{graphicx}                
  \DeclareGraphicsExtensions{.pdf,.png,.jpg,.jpeg} 
\else
  \usepackage{graphicx}                
  \DeclareGraphicsExtensions{.eps}     
\fi%

\graphicspath{{figures/}{pictures/}{images/}{./}} 

\usepackage{microtype}                 
\PassOptionsToPackage{warn}{textcomp}  
\usepackage{textcomp}                  
\usepackage{mathptmx}                  
\usepackage{times}                     
\usepackage{cite}                      
\usepackage{tabu}                      
\usepackage{booktabs}                  

\onlineid{0}

\vgtccategory{Research}

\vgtcinsertpkg

\preprinttext{This manuscript was  presented at alt.VIS, a workshop co-located with IEEE VIS 2021 (held virtually)}

\title{Touching Art - A Method for Visualizing Tactile Experience}

\author{Bernice Rogowitz\thanks{e-mail: bernice.e.rogowitz@gmail.com}\\ %
    \scriptsize Visual Perspectives Research %
\and Laura J. Perovich\thanks{e-mail: l.perovich@northeastern.edu}\\ %
    \scriptsize Northeastern University %
\and Yuke Li\thanks{e-mail: li.yuke1@northeastern.edu}\\ %
    \scriptsize Northeastern University %
\and Bjorn Kierulf\thanks{e-mail: kierulf.b@northeastern.edu}\\ %
    \scriptsize Northeastern University %
\and Dietmar Offenhuber\thanks{e-mail: d.offenhuber@northeastern.edu}\\ %
    \scriptsize Northeastern University %
}

\teaser{
  \centering
  \includegraphics[width=\linewidth]{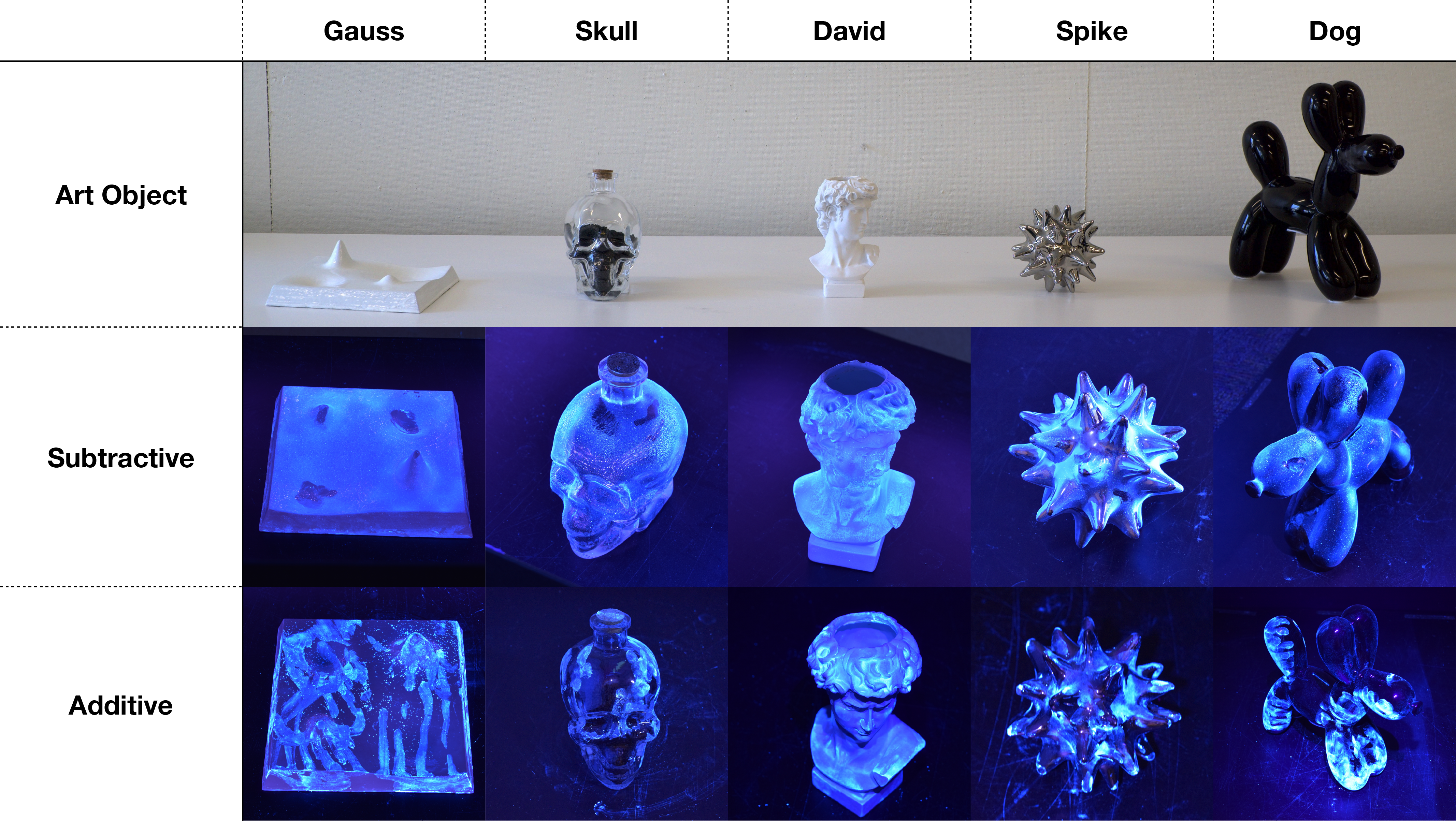}
  \caption{Three-dimensional art objects and their touch patterns, revealed by fluorescent imaging.  Results for the subtractive and additive method are shown. These methods reveal a range of touching, tracing, and grasping gestures, which are useful for exploring how humans interact with art and real-world objects.}
  \label{fig:teaser}
}

\abstract{
It is human to want to touch artworks, to feel their surface curvature and texture, their shapes and structures, and to feel the hand of the artist \cite{candlin2006}. Museum guards need to be constantly vigilant to protect art objects from adoring and exploring touches by visitors. This paper introduces a novel technique for capturing where and how art objects are touched. In this method, the users' touch either adds, or subtracts, microscopic fluorescent particles from a three-dimensional art object. Viewing the object under ultraviolet light reveals their touch traces and gestures. We present human touch behavior for a three-dimensional stylized landscape, and for two abstract and two representational art objects.  We also present the results of video recordings of real-time behavior and user interviews. The resulting data show the kinds of touches, and where they are directed, and also reveal important individual differences. We feel this method opens the door to studying art perception through touch, and also enables new kinds of studies into touch behavior in other applications, including visualization, embodied cognition, and design.  
} 

\CCScatlist{
  \CCScatTwelve{Human-centered computing}{Visu\-al\-iza\-tion}{Visu\-al\-iza\-tion techniques};
  \CCScatTwelve{Human-centered computing}{Visu\-al\-iza\-tion}{Visualization design and evaluation methods}{Haptics}
}

\begin{document}


\firstsection{Introduction}

\maketitle

Touch is a natural way people interact with objects. In her article called “Art, Museums and Touch,” Fiona Candlin describes how touching artworks gives immediate insight into how a surface is explored and understood.  It lets people push boundaries, such as publicly patting the head of a lion sculpture or touching Juliette’s breast, and the way people touch, through points and strokes, and caresses, reveals emotional reactions \cite{candlin2006,candlin2009}. Capturing these touches on three-dimensional objects, however, is a technical challenge. Direct touch methods, such as pressure sensitive sensors, require significant instrumentation (e.g., piezoelectrics), often require modifying the object to accept sensors, and may require the user to wear sensor gloves that interfere with the touch behavior itself. Moreover, technological sensing approaches are costly and require bespoke solutions for different objects. Video methods can non-intrusively approximate where an object has been touched, but cannot reliably capture the variety of touches and gestures. 

This paper describes the first application of a novel methodology we are developing for non-invasively capturing a wealth of human touches on two- and three-dimensional objects. In this method, we use a tracer substance called \textit{GloGerm}\footnote{See \url{https://www.glogerm.com}}, which is used to simulate how germs are spread by hand in healthcare and food processing applications. In subtractive touch, objects are sprayed with a thin, nearly-invisible powder coating of these microscopic ultraviolet fluorescent \textit{GloGerm} particles. When the object is touched, the powder is removed, and the touched regions do not fluoresce under ultraviolet light.  In the additive method, this fluorescent powder is applied to the observer’s hands, and is transferred to the object’s surface when touched. In this scenario, the touched regions fluoresce under ultraviolet light. Our upcoming paper establishes the accuracy and reliability of these methods to capture touch and trace gestures, and shows how they can be used to capture touch behaviors associated with directed tasks on constructed surfaces and objects. In this paper, we explore human tactile behavior on more complex, artistic, objects, to reveal what we touch when we are allowed to touch art objects. We discuss the range of motions and touches used in a free-touch environment, and explore individual differences. The ability to measure where and how people touch art objects informs our understanding of how we interact with art, but also provides a methodology for studying tactile exploration in other contexts.

\begin{figure}
    \centering
    \includegraphics[width=\linewidth]{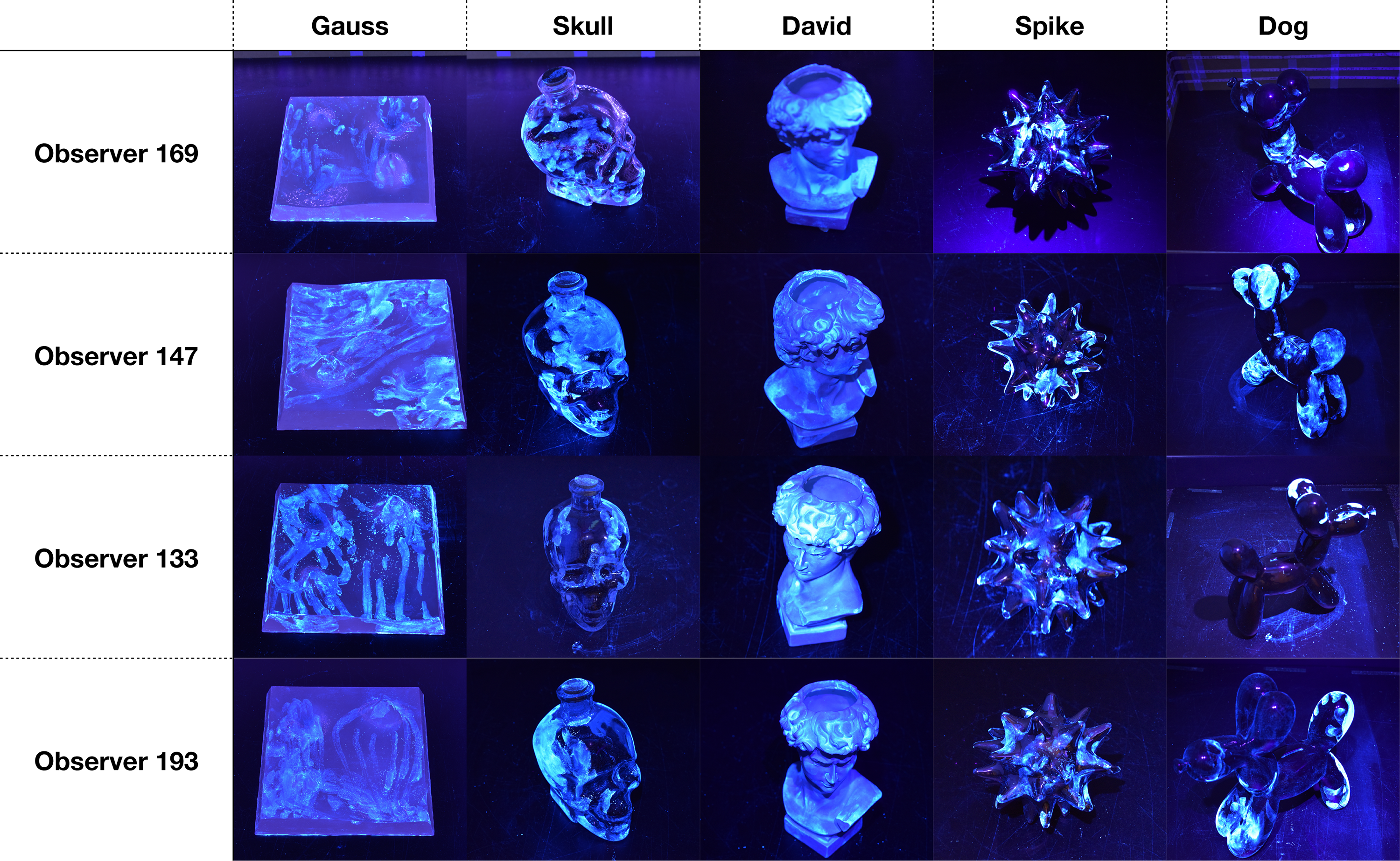}
    \caption{Touch patterns on art objects using the additive method}
    \label{fig:add}
\end{figure}

\section{Touch in visualization research} 
An additional motivation for this study is to break ground in a largely unexplored area of visualization and physicalization research: the role of touch for making sense of data representations. Research in psychology offers many starting points: people learn to interpret charts by tracing them with their fingers even if they don’t offer tactile features \cite{agostinho2015}, touching objects often supports learning and memorizing their features better than just observing them \cite{novak2020,hutmacher2018}. Tactile stimulation is used extensively in the Montessori pedagogy; letters cut from sandpaper help children to learn the alphabet with multiple sensory modalities \cite{novak2020}. Also data physicalization research has shown that the ability to touch a tangible data object improves its understanding \cite{jansen2013}. So far, however, the visualization discipline has largely overlooked the touch modality. Some data physicalization researchers have used haptic feedback to represent data, though often this involves single-channel touch information such as temperature or vibration \cite{hogan2013,brueckner2018}.  Research on the design of tactile graphics and data visualizations for the visually impaired has led to many new insights \cite{edman1992,engel2021}, though these results may not generalize to the larger population, since reading tactile charts requires specialized skills that individuals without visual impairments typically do not acquire. 

\section{Human touch behaviors and object qualities}
Attempts at generalization are limited by the finding that touch behaviors are individual-specific. “Need for touch” (NFT) scales are one way to capture and characterize these behaviors along two dimensions: \textit{instrumental NFT} and \textit{autotelic NFT} \cite{peck2003}. People high in instrumental NFT use touch to gather information and make judgments about objects. Those high in autotelic NFT are hedonically motivated to touch objects and enjoy the sensory and exploratory aspects of touch \cite{peck2003}. Research using images of objects has shown broad human preferences for touch related to qualities of the target object. Rough surfaces are often perceived as less touchable — for example, glass is preferred to concrete — though very smooth surfaces may also be perceived as less touchable \cite{klatzky2011}. Very simple and very complex object shapes are often perceived as less touchable \cite{klatzky2011}, as well as less aesthetically pleasing \cite{spehar2003universal}. Stroking touch behaviors are particularly related to the perceived touchability of objects and texture appears to be a primary driver of these behaviors \cite{klatzky2011}. 

\section{Interpretation of traces}
Our tracer substance generates persistent marks that offer a compelling record of the user’s interaction with the object — a form of autographic visualization \cite{offenhuber2019}. Beyond the basic question of where the object is touched, the traces offer clues about the quality of touch: which parts of the hand are used, the intensity of touch, whether fingers are moved across the surface. As unobtrusive measures, traces can offer a rich and multi-faceted picture and can be interpreted from many different angles \cite{webb1979}. Webb differentiates between traces of erosion and of accretion \cite{webb1966}, which are also popular design metaphors in HCI and visualization, to convey ideas of temporality and accumulation \cite{huron2013,hill1992}. In the context of our study, these correspond to the subtractive and additive methods. We used our method to capture touch and trace gestures on art objects, in a controlled setting. To help us begin interpreting the meaning behind these touches, we also recorded videos of the interaction and conducted semi-structured interviews. Our goal is to provide tools to explore the semantics of touch.  

\section{Methods and study design}
These experiments were designed to test the use of our subtractive and additive fluorescent imaging methods to capture the many ways art objects are touched. To simulate the museum experience, the objects were set out on a table for display. The observers were free to touch the objects as they chose, not driven by a specific task.  

\subsection{Stimuli}
We selected five objects for study, which are shown in the first row of \autoref{fig:teaser}. These included a simulated landscape, two abstract non-representational objects, and two representational objects. In the order shown in \autoref{fig:teaser} from left to right, the objects were:

\begin{enumerate}
\item “\textbf{Gauss}” Gaussian surface sculpture created by overlaying four positive and negative Gaussian functions. A rectangular object milled from medium density fiber (MDF) board, and coated with a semi-rigid, white opaque layer of epoxy resin, which appears smooth, although imperfections of the coating are visible and tangible.
\item “\textbf{Skull}” Novelty glass decanter that contains multiple contours representing features of the human skull with a bottle neck and a cork on top. It is a smooth and transparent glass object that was filled with black lentils to increase visual contrast.
\item “\textbf{David}” Classic bust, inspired by Michelangelo’s David. This reproduction is made from thin resin and has a rough surface.  
\item “\textbf{Spike}” Spherical ceramic object covered with rounded 1-inch high convex spikes. Its smooth mirror surface appears to be made from metal, but is actually a ceramic glazing.
\item “\textbf{Dog}” Novelty bank reproduction of the famous balloon dogs created by Jeff Koons. This reproduction is a hollow ceramic object with double curved surfaces and smooth black glazing, with a coin slot and removable plastic stopper.  
\end{enumerate}

\begin{figure}
    \centering
    \includegraphics[width=\linewidth]{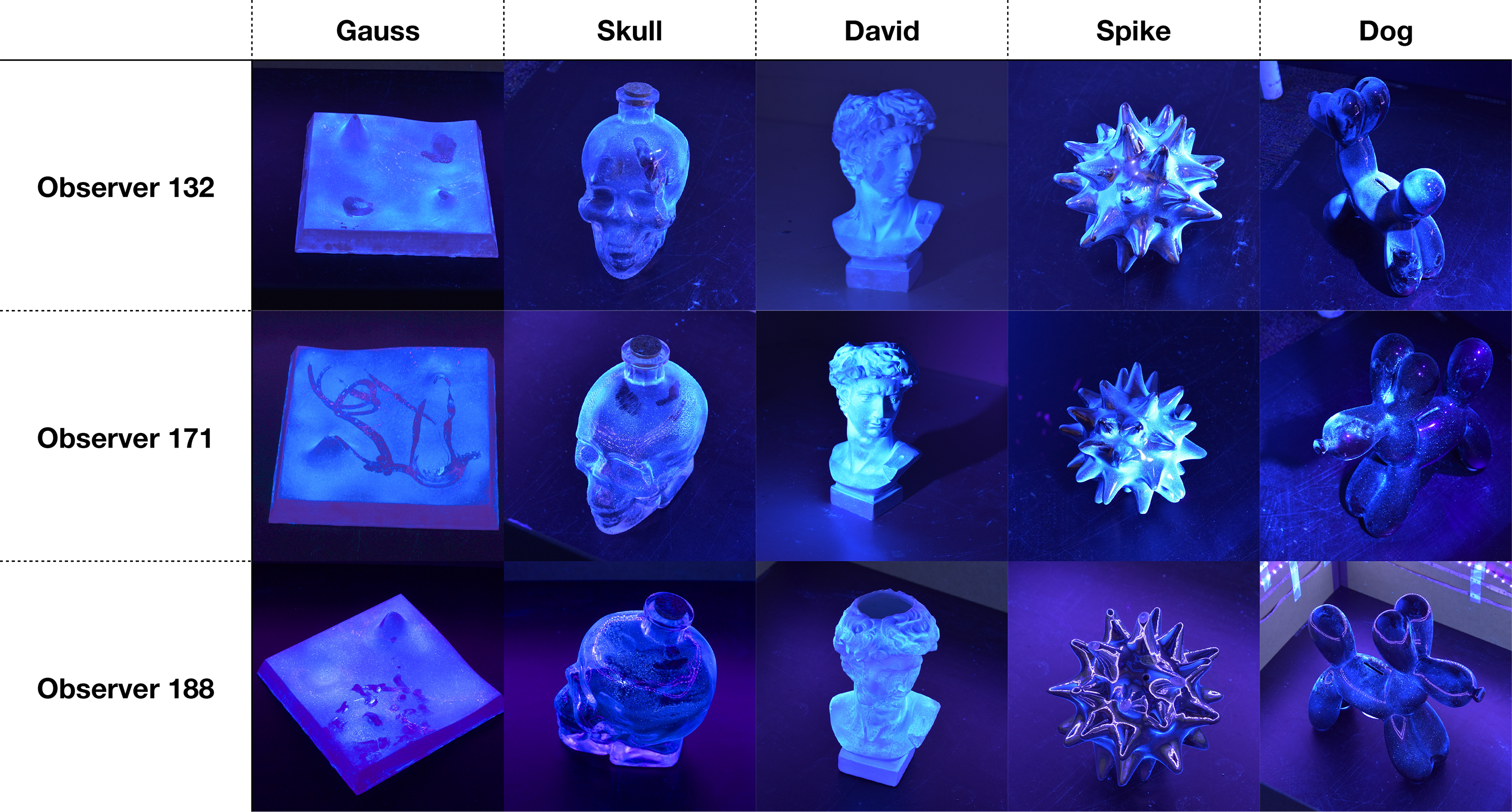}
    \caption{Touch patterns on art objects using the subtractive method.  }
    \label{fig:sub}
\end{figure}

\subsection{Experimental Design} 
 Data were collected in a blocked design, with different observers serving in the additive and subtractive conditions. For both conditions, the five objects were arrayed on a table, in varying orders. In the subtractive condition, the \textit{GloGerm} powder was suspended in alcohol and air-brushed onto the object. This produced a nearly-invisible layer of fluorescent particles. In the additive condition, the fluorescent powder was applied to the observers' hands before touching the objects. In both conditions, the observer was instructed to examine the objects, and to feel free to touch them however they wanted. After each session, we photographed the objects under ultraviolet light, from multiple directions, to capture an objective measure of their touch patterns (\autoref{fig:teaser}). To capture the subjective experience of our observers, we video recorded their movements. We used the video to better understand observers' exploration process, to contextualize the photographed touch outcomes with the touch behaviors, and to capture incidental comments as they conducted the task. At the end of each session, we conducted a semi-structured interview, where we asked the observer what the experience was like for them, which objects were their favorite, and if the objects felt as they expected. The fluorescent material was removed from the objects between sessions.

Three subjects participated in the subtractive condition, two identifying as male, one identifying as female. Four subjects, all identifying as female, participated in the additive condition. Experiments took place in July 2021. Participants were in their 20s and 30s.

\section{Results}
\subsection{Fluorescent Images of Observer Touches}
The subtraction and addition of the fluorescent powder provided an objective measure of where the objects were touched, which we captured by photographing the objects under ultraviolet light. \autoref{fig:add} shows results from four subjects using the additive method. \autoref{fig:sub} shows the results from four subjects using the subtractive method. Both methods produced reliable results and demonstrated touches and gestures made by a single finger, multiple fingers, and the whole hand. Both methods produced clear results when viewed under the ultraviolet light by the human eye. The photographic images produced when the fluorescing material was applied to the object (additive method) were more effective in capturing these touch patterns. In the additive method, more fluorescing material was present, producing higher contrast photographs. In the subtractive method, less powder was used and in some cases residual powder remained on the objects after cleanings which reduced the contrast of the images. 

\subsection{Types of Touches} 
\begin{figure}[b]
    \centering
    \includegraphics[width=\linewidth]{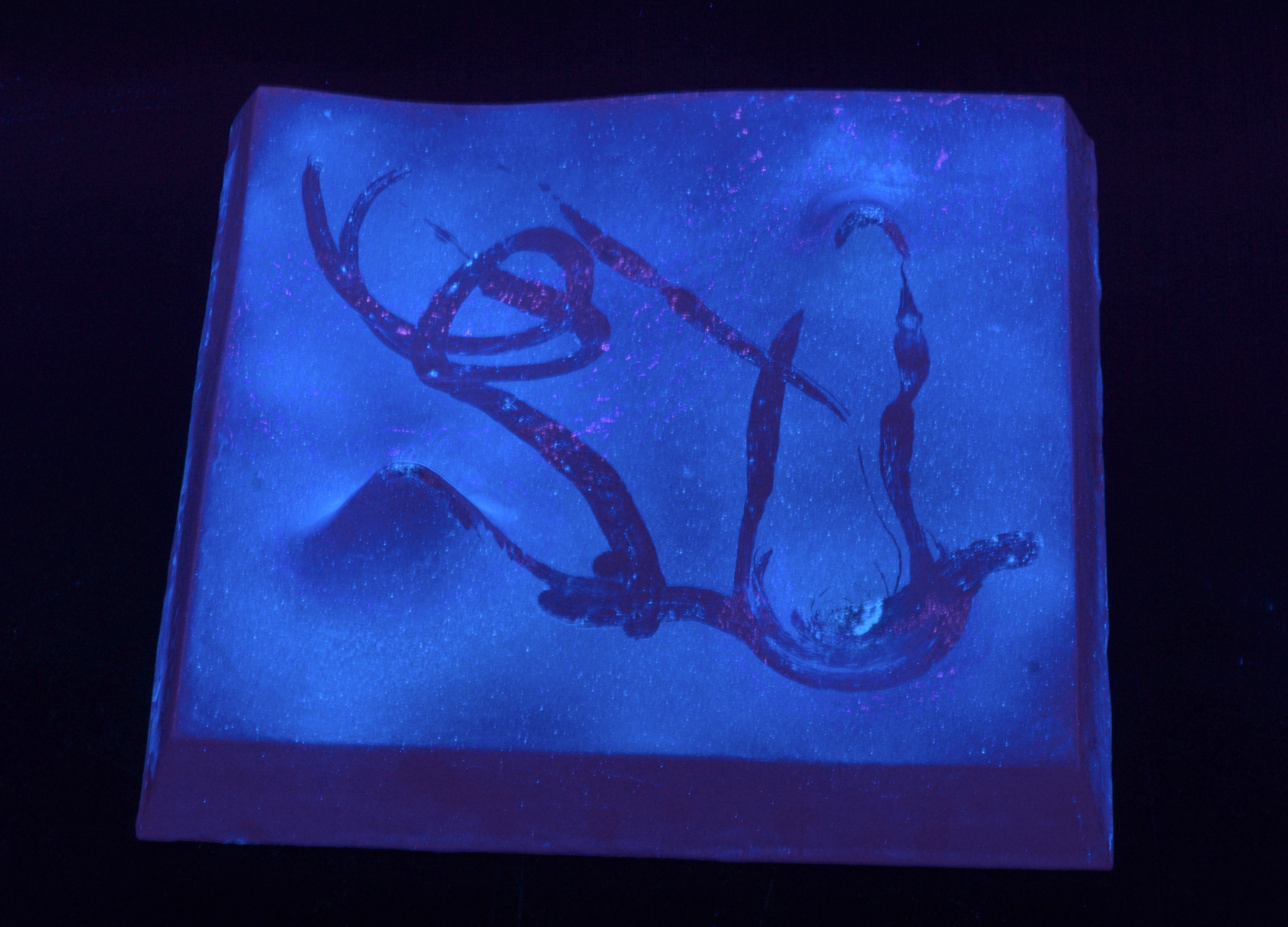}
    \caption{Subject 171, subtractive traces on the object "Gauss."  }
    \label{fig:gauss}
\end{figure}

\begin{figure}
    \centering
    \includegraphics[width=\linewidth]{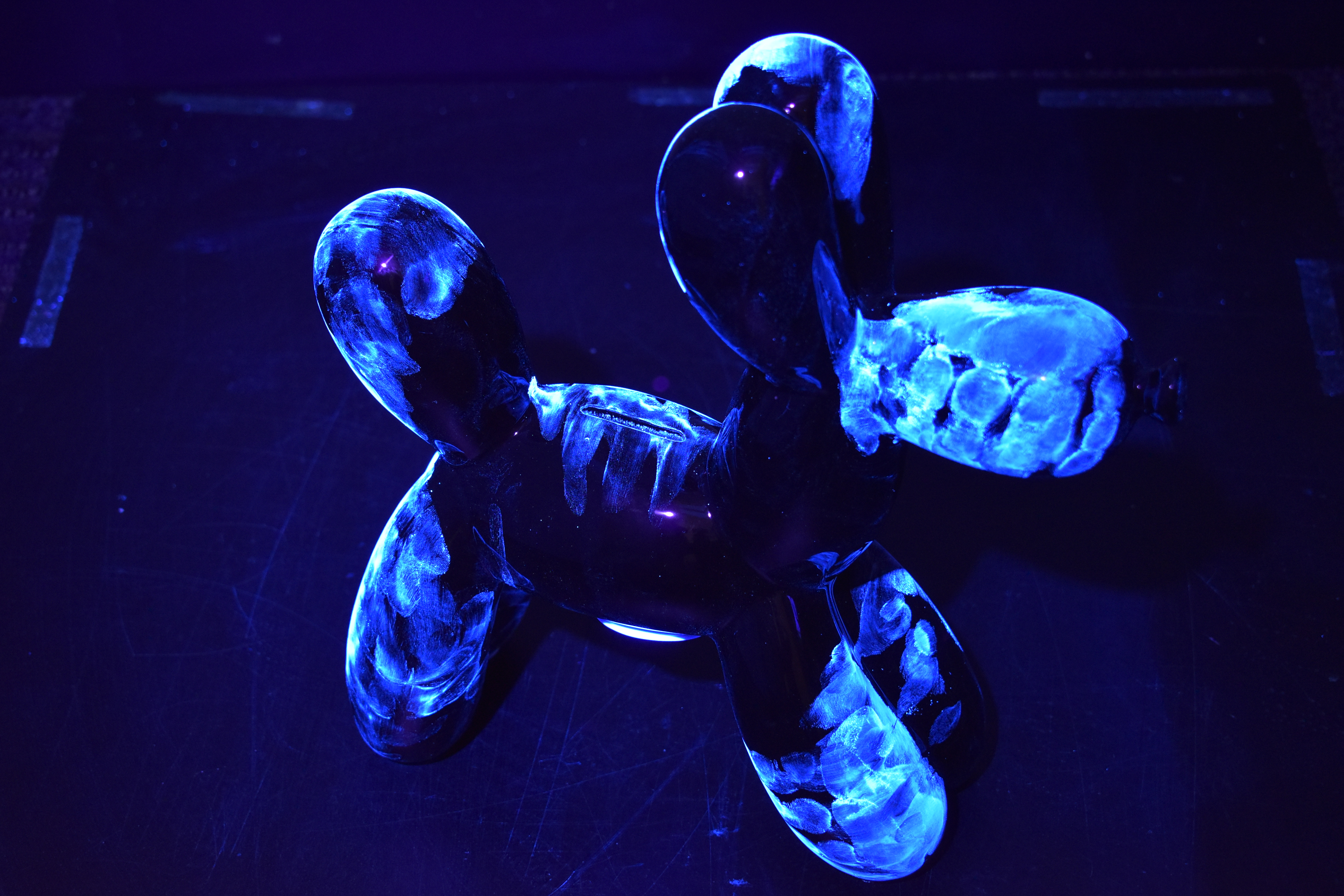}
    \caption{Subject 147, additive traces on the object Dog.  }
    \label{fig:dog}
\end{figure}

For each observer, we examined the photographic images of their touch patterns and compared these with the video recordings of their touching behavior. For both methods, we observed individual regions where the powder had been deposited or removed, which we identified as touches or path tracing by a single finger. Figure \ref{fig:gauss} shows a clear pattern of complex tracings on and around the peaks of Gauss. We also observed whole-hand grasping patterns, where the imprint of several fingers was captured. Figure \ref{fig:dog} shows how the legs and tail of Dog were grasped. Large regions where the material was either added or removed indicated places where the shape or curvature of the object was examined. 


\subsection{Relating Touch Images to Touching Behaviors}
To understand images of touch patterns more fully, we compared them to the touching behaviors captured in the video. In the videos, we observed (1) one finger or multi-finger pointing and stroking gestures, including path tracing, (2) whole hand stroking gestures, including feeling textures and contours, (3) multi-finger grasping gestures, including pinching objects to rotate them while they remain on the table, (4) whole hand grasping gestures, including holding, lifting, and enveloping objects in one or two hands. These touch patterns revealed a wide range of different exploratory strategies. We observed that for most objects, the first step for all participants was to lift the object. When lifting objects, some participants used one hand and turned the object around to look at it from all sides. Others lifted it with two hands and rotated it in their palms to change orientation. They cupped or grasped particular parts of the object, and when they explored a specific region, they used the thumbs and the sides of their fingers for tracing, not just the fingertips. 

We made some preliminary observations on the types of touches used on various objects. We noticed that Gauss was not lifted as frequently as the other objects and that it elicited more tracing behaviors, especially on the peaks and concavities. Participants used their fingers to trace paths and points of interest, used multiple fingers along curved surfaces, and also used their whole hand to trace over the object. Many observers grasped the tail or nose of Dog with their whole hand and often lifted it, using these areas as handles. Spike was often held in one hand with the fingers between the spikes and Skull was often held in two hands. The miniature David attracted fewer touches than the other objects. 

\subsection{Additive versus Subtractive}
The additive  and the subtractive methods have their advantages for different objects and for emphasizing different aspects of the interaction.  The traces captured by the subtractive method had high precision, but tended to be light and sometimes hard to capture photographically (see \autoref{fig:sub}). The associated videos show that the subjects were more tentative in exploring objects that were prepared in this way. This was unexpected: with the powder almost invisible, the subtractive method was less intrusive. The additive method produced more robust, higher-contrast traces. The participants had to dip their hands into the powder, which was initially unpleasant for some, but once that hurdle was overcome, their touches were less tentative than with the subtractive method.   

Both methods produced artifacts that limit their applicability. In case of the subtractive method, it can be difficult to create an even powder coating, and uneven areas can be confused with traces. For the additive method, it makes a difference whether participants thoroughly cover their entire hands in powder.  

\subsection{Tracing and Grasping}
 The majority of the impressions we captured reflected overall holding, moving, and turning the object with the whole hand. Gauss elicited the most targeted touches and tracing gestures, perhaps because it was heavier, and more awkward to pick up.  Many single-finger traces on Gauss were reminiscent of visual scan paths, captured by eye tracking, and to how visually impaired people are trained to explore tactile information.  We also observed multiple-finger traces along contours and whole-hand impressions.  We observed an interesting interplay between visual inspection, kinesthetic examination (such as shaking the object), and tactile exploration. Many senses are used at the same time and support each other. With regard to tactile exploration, we also observed an interplay between the interrelated tasks of holding and supporting the object and exploring its surface. To explore tactile features, the participants  not only used their fingertips, but also their thumbs, the sides of their fingers, their palms, and the whole hand. 

\subsection{Individual Differences Between Observers}
All observers, in both conditions, touched all the objects. However, individuals varied in their approach. Participants split their time differently between the objects and occasionally touched the objects in a different order than was suggested by their linear arrangement on the table. Some participants spent more time touching objects while others spent more time observing them visually. We  observed differences in the style of touching. Some observers were tentative in their touches, others were enthusiastic and autotelic (S147), taking aesthetic pleasure in the touch process. Some seemed analytical, tracing contours of the object (S171), others seemed to explore and caress the objects with painterly gestures (S193) and others simply picked them up, turned them around and moved on. Gauss produced the greatest variation in responses between observers. 

\subsection{Participant Observations}
The participants commented on their experience during and after interacting with the objects. For some objects, they mentioned that touching gave them information that conflicted with their visual understanding. The apparently metallic, but actually ceramic, material of Spike was surprising (S132) and Dog was described as heavier than expected (S193). Curiosity about the material and particular features of the objects were cited as motivating their touch (S133). Some participants noted a change in interest in a particular object after touching it, finding it better than expected (S133). Another observer found the indentations of Gauss and the smooth surface of Dog to be attractive and appreciated how easy it was to handle Dog (S169). One participant expressed discomfort touching Spike because it was unclear how to hold it. Participants also talked about their past experience in interacting with the objects; for example, one participant with a physics background mimicked the path of a marble on Gauss (S171) and another mentioned disliking an object because of bad memories of drawing classes (S147). 

Subject differed in their appreciation of the powder. At least one participant enjoyed dipping their hands into the tracer substance, remarking that the powder made their hands less sweaty and gave them more confidence in judging the object materials (S193).  

A few people used sound to explore the objects. One remarked, “I knock it because I want to distinguish are they all made by the same material. I can distinguish those by the sound” (S193). Others were observed tapping or shaking the objects.


\section{Discussion}

\begin{figure}
    \centering
    \includegraphics[width=\linewidth]{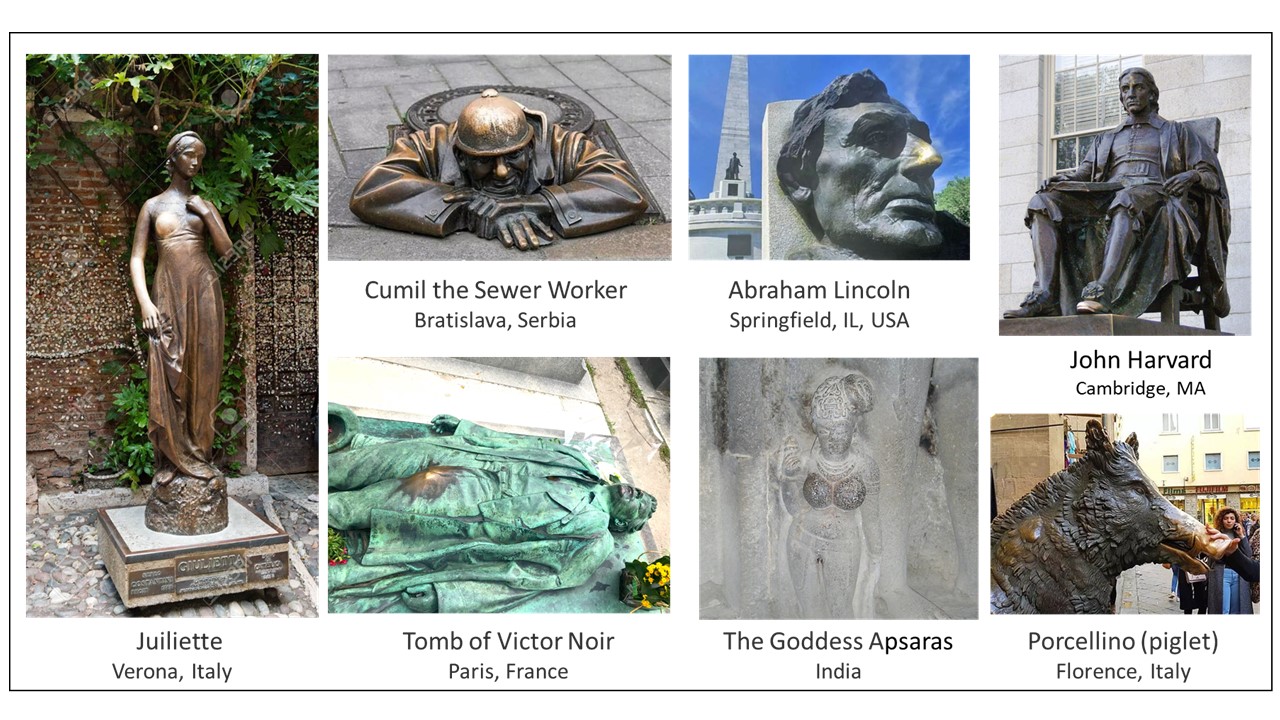}
    \caption{Examples of autotelic responses, where we are drawn to touch art objects from around the world.  }
    \label{fig:auto}
\end{figure}

Figure \ref{fig:auto} shows sculptures from around the world that attract human touch. Over the decades, people have been drawn to touch a nose, a shoe, or another feature. These autotelic touches have been captured by the gradual polishing of the bronze or the gradual smoothing of the stone.  Our experiments aim to develop a method that likewise captures where and how we touch art objects, but on a much shorter time scale.

In our method, microscopic fluorescent particles are either added or removed from the surface when touched. Viewing these additions or subtractions under ultraviolet light provides a clear record of where the object was touched. Using this method, we were able to capture and record different types of touch. In some cases, we were able to distinguish which part of the hand created the mark, and whether the object was actively explored or just held. Despite the limited number of participants, the collected data show a surprisingly rich variety of tactile interactions with the objects. We also created video recordings of our observers' touch behaviors, as a first step toward interpreting the meaning of these touches. We found that our observers not only touched the objects, traced paths on their surfaces, caressed their contours and pinched their peaks, they often picked them up, rotated, and explored them. In all cases, our method captured the touch behavior, but interpreting this behavior is quite complex. Different objects elicited different touch behaviors, specific features of the object elicited different touch behaviors, and individuals displayed different touch styles.

\subsection{Touching Physical Objects}
We noticed that participants were more interested in exploring the larger, heavier objects, Gauss and Dog, than the smaller, lighter ones. In the videos, it seemed that the smaller objects were treated like items in a gift shop. They were picked up and examined, often cursorily, and sometimes only visually, and didn't attract the variety of traces and strokes the more substantial objects received. Gauss elicited more exploratory and differentiated touches. The limbs of Dog were large enough to fit perfectly into the palm of the adult hand and seemed to invite this kind of grasping. While large-scale sculptures, such as the Juliette in Verona and Michelangelo's David, encourage touch, it seems that their small-scale replicas may not be as compelling. For example, our observers showed particularly little interest in the head of David replica, which was quite small, and made of a very light resin. In the future, we would like to explore how scale affects how an object is touched.
Variations in how participants touch particular objects may be partially caused by other object qualities. We saw that single finger or whole hand tracing gestures were common on Gauss, which aligns with Klatzky et al's finding that smooth objects with medium complexity are most inviting to touch, especially with stroking gestures \cite{klatzky2013,klatzky2011}. Perhaps it wasn't the size of the David sculpture that reduced its attraction, but its complexity or rougher surface. However, Klatzky's results were obtained by subjects viewing pictures of objects. Future work could repeat those experiments with real-world objects combined with our method for recording touch behavior.   


This method could also be used to explore the affordances of objects more systematically, for example, whether objects that seem graspable are in fact grasped more often. Also, our objects were all static and stationary. This technique could be very useful to study how people interact with composite objects with visible joints or parts that can be moved or dismantled.

It would also be interesting to further explore the tactile appreciation of artworks. Having a lightweight non-invasive method could provide insights into how the physical characteristics of a sculpture or art object encourages different types of touches, how the conceptual meaning of the piece changes those touch patterns, and how characteristics of the people who touch them play in this equation.  

\subsection{Documentation of the Traces}
Although we were able to capture touch impressions photographically, improvements in our method may yield better results. We found that seeing and interpreting touch traces was much easier when actively inspecting the objects under ultraviolet light than when viewing the photographic images. This may be because viewing a three-dimensional object directly provides more information than we can capture in two-dimensional images of three-dimensional objects. We also found that many precautions were required to carefully calibrate the contrast of the traces against the background and minimize specular reflections in the photographs. 

\subsection{Implications for Visualization}
Data visualization research is intimately connected with the study of visual perception. Visualization design is both guided and constrained by what our visual system can discriminate and recognize,
and consequently many researchers engage in experimental studies of perception. Data physicalization extends data representations into physical space and consequently requires the study of tactile and multi-sensory perception. At the moment, research here remains limited and many data physicalizations are designed for visual consumption. In some cases, this makes sense as many visualizations can be readily translated into objects. However, vision and touch provide different channels of information. Methods that allow us to learn more about how we acquire and process information through touch, and how we use touch to analyze and explore physical data representations, can provide deep insight into these processes.  

The capacity to touch could also be important for interaction research, such as understanding the interaction with building maps or architectural models that are sometimes used as navigational aids in complex interior environments.  It could also be useful to study the process of skill acquisition for using manual tools and interacting with unfamiliar interfaces. Future variations of this method, using different tracer powders for different users, could be useful for studying the collaborative exploration of objects.
While we opted for a more structured environment, our approach is extensible to other real-world environments where the response is not constrained, allowing behavioral experiments in the wild. 
\section{Conclusion}
We have demonstrated a novel technique that can be used to visualize how people touch art objects. Both the additive and subtractive methods captured the observers' touches. In these experiments, the additive method seemed to encourage more expressive gestures and resulted in higher contrast images; the subtractive method allowed us to capture more nuanced touches and gestures. The two methods, thus, might be best tuned for different touch tasks. For example, if the goal is to capture the emotional response to sculptures, the additive method may be better suited; if the goal is to capture gestures for exploring the fine-structure of a 3-D visualization of scalar data, the subtractive method might be more appropriate.  

The ability to reliably capture touch behavior can be used in a wide range of different application areas. It can support perceptual research aimed at understanding how touch patterns depend on object characteristics, such as roughness and complexity.  It could also be used as a tool for data physicalization, for example, to measure how tactile exploration varies depending on the task, or to support research in embodied cognition, and in other domains where knowing where and how people touch objects can inform our understanding of the thought processes underlying these touches.

Being able to reliably measure where and how people use touch is a key first step toward addressing the semantics and intent behind touch gestures. In this experiment, video recording the observers while they touched the art objects and capturing verbal feedback on their interactions provided some preliminary insight. We think this is an important research direction, which can open the door to understanding touch behavior and its role in cognitive processing. 


\acknowledgments{We thank the organizers of the 2018 Dagstuhl workshop on Data Physicalization for launching this dialog between the authors, the Northeastern University College of Arts, Media and Design Cluster Collaborative Seed Grant and the Summer Student Support Grant, and Patrick Kana for fabrication support. Thanks to Gordon Rogowitz for his photograph of The Goddess Apsaras in Figure 6.
}

\bibliographystyle{abbrv-doi}
\bibliography{touch}

\begin{thebibliography}{10}

\bibitem{agostinho2015}
S.~Agostinho, S.~Tindall-Ford, P.~Ginns, S.~J. Howard, W.~Leahy, and F.~Paas.
\newblock Giving {{Learning}} a {{Helping Hand}}: {{Finger Tracing}} of
  {{Temperature Graphs}} on an {{iPad}}.
\newblock {\em Educational Psychology Review}, 27(3):427--443, 2015-09. doi:
  {{%
10\hspace{.1pt}\discretionary{.}{%
}{.}\hspace{.4pt}1007\discretionary{/}{%
}{/}s10648\discretionary{%
}{-}{-}015\discretionary{%
}{-}{-}9315\discretionary{%
}{-}{-}5}}


\bibitem{brueckner2018}
S.~Brueckner and R.~Freire.
\newblock Embodisuit: A wearable platform for embodied knowledge.
\newblock In {\em Proceedings of the {{Twelfth International Conference}} on
  {{Tangible}}, {{Embedded}}, and {{Embodied Interaction}}}, pp. 542--548,
  2018.

\bibitem{candlin2006}
F.~Candlin.
\newblock The dubious inheritance of touch: {{Art}} history and museum access.
\newblock {\em Journal of Visual Culture}, 5(2):137--154, 2006.

\bibitem{candlin2009}
F.~Candlin.
\newblock {\em Art, Museums and Touch}.
\newblock {Manchester University Press}, 2009.

\bibitem{edman1992}
P.~Edman.
\newblock {\em Tactile Graphics}.
\newblock {American Foundation for the Blind}, 1992.

\bibitem{engel2021}
C.~Engel, E.~F. Müller, and G.~Weber.
\newblock Tactile {{Heatmaps}}: {{A Novel Visualisation Technique}} for {{Data
  Analysis}} with {{Tactile Charts}}.
\newblock In {\em The 14th {{PErvasive Technologies Related}} to {{Assistive
  Environments Conference}}}, pp. 16--25, 2021.

\bibitem{hill1992}
W.~C. Hill, J.~D. Hollan, D.~Wroblewski, and T.~McCandless.
\newblock Edit wear and read wear.
\newblock In {\em Chi}, vol.~92, pp. 3--7, 1992.

\bibitem{hogan2013}
T.~Hogan and E.~Hornecker.
\newblock In touch with space: Embodying live data for tangible interaction.
\newblock In {\em Proceedings of the 7th International Conference on Tangible,
  Embedded and Embodied Interaction}, pp. 275--278, 2013.

\bibitem{huron2013}
S.~Huron, R.~Vuillemot, and J.-D. Fekete.
\newblock Visual sedimentation.
\newblock {\em IEEE Transactions on Visualization and Computer Graphics},
  19(12):2446--2455, 2013.

\bibitem{hutmacher2018}
F.~Hutmacher and C.~Kuhbandner.
\newblock Long-{{Term Memory}} for {{Haptically Explored Objects}}:
  {{Fidelity}}, {{Durability}}, {{Incidental Encoding}}, and {{Cross}}-{{Modal
  Transfer}}.
\newblock {\em Psychological Science}, 29(12):2031--2038, 2018-12-01. doi: {{%
10\discretionary{/}{%
}{/}gfssc8}}


\bibitem{jansen2013}
Y.~Jansen, P.~Dragicevic, and J.-D. Fekete.
\newblock Evaluating the efficiency of physical visualizations.
\newblock In {\em Proceedings of the {{SIGCHI Conference}} on {{Human Factors}}
  in {{Computing Systems}}}, pp. 2593--2602. {ACM}, 2013. doi: {{%
10\discretionary{/}{%
}{/}gf65hg}}


\bibitem{klatzky2013}
R.~L. Klatzky and S.~J. Lederman.
\newblock Touch.
\newblock In A.~Healy and R.~Proctor, eds., {\em Handbook of Psychology:
  {{Experimental}} Psychology}, pp. 147--176. {John Wiley \& Sons Inc.}, 2013.

\bibitem{klatzky2011}
R.~L. Klatzky and J.~Peck.
\newblock Please touch: {{Object}} properties that invite touch.
\newblock {\em IEEE Transactions on Haptics}, 5(2):139--147, 2011.

\bibitem{novak2020}
M.~Novak and S.~Schwan.
\newblock Does {{Touching Real Objects Affect Learning}}?
\newblock {\em Educational Psychology Review}, 2020-08-19. doi: {{%
10\discretionary{/}{%
}{/}ghk9x7}}


\bibitem{offenhuber2019}
D.~Offenhuber.
\newblock Data by {{Proxy}} \textemdash{} {{Material Traces}} as {{Autographic
  Visualizations}}.
\newblock {\em IEEE transactions on visualization and computer graphics}, 2019.
  doi: {{%
10\discretionary{/}{%
}{/}gf649f}}


\bibitem{peck2003}
J.~Peck and T.~L. Childers.
\newblock Individual differences in haptic information processing: {{The}}
  “need for touch” scale.
\newblock {\em Journal of Consumer Research}, 30(3):430--442, 2003.

\bibitem{spehar2003universal}
B.~Spehar, C.~W. Clifford, B.~R. Newell, and R.~P. Taylor.
\newblock Universal aesthetic of fractals.
\newblock {\em Computers \& Graphics}, 27(5):813--820, 2003.

\bibitem{webb1979}
E.~Webb and K.~E. Weick.
\newblock Unobtrusive {{Measures}} in {{Organizational Theory}}: {{A
  Reminder}}.
\newblock {\em Administrative Science Quarterly}, 24(4):650, 1979-12. doi: {{%
10\discretionary{/}{%
}{/}chnvg9}}


\bibitem{webb1966}
E.~J. Webb, D.~T. Campbell, R.~D. Schwartz, and L.~Sechrest.
\newblock {\em Unobtrusive Measures}.
\newblock {Sage Publications}, 1966.

\end{thebibliography}
\end{document}